\documentclass[prb,twocolumn]{revtex4}
\usepackage{epsfig}
\usepackage{graphicx}
\usepackage{amsfonts,amssymb}
\usepackage{amsmath}
\usepackage{color}
\usepackage {times}

\usepackage{bm}
\usepackage{blindtext}
\usepackage{sidecap}
\usepackage{floatrow}

\usepackage[utf8]{inputenc}
\usepackage[T1]{fontenc}
\usepackage{mathptmx} 
\definecolor{refcolor}{rgb}{1.0,0.0,0.0}

\newcommand{\be}{\begin{equation}}
\newcommand{\ee}{\end{equation}}   
\newcommand{\bea}{\begin{eqnarray}}
\newcommand{\eea}{\end{eqnarray}}
\newcommand{\ba}{\begin{array}}
\newcommand{\ea}{\end{array}}

\begin{document}

\title{Hund's coupling and electronic anisotropy in the spin-density wave state of iron pnictides}

\author{Garima Goyal and Dheeraj Kumar Singh}

\affiliation{School of Physics and Materials Science, Thapar Institute of Engineering and Technology, Patiala-147004, Punjab,India}     

\begin{abstract}
In the multiband systems, Hund's coupling ($J$) plays a significant role in the spin and charge excitations. We study the dependence of electronic anisotropy on $J$ in terms of Drude-weight along different directions as well as the orbital order in the four-fold symmetry broken spin-density wave state of iron pnictides. A robust behavior of the Drude-weight anisotropy within a small window around $J \sim 0.25U$ with $U$ as intraorbital Coulomb interaction is described in terms of orbital-weight distribution along the reconstructed Fermi surfaces. We also find that the ferro-orbital order increases with $J$ in the widely accepted regime for the latter, which is explained as a consequence of rising exchange field with an increase in magnetization.
\end{abstract}

\maketitle
\section{Introduction}
Parent compounds of unconventional iron-based superconductors exhibits
a collinear spin-density wave (SDW) order which is metallic~\cite{lee, johnston,stewart, dai}
unlike their high-$T_c$ cuprate counterparts. The SDW state exhibits highly
anisotropic electronic properties~\cite{shimojima, greve,liang, fernandes, fernandes1, blomberg,kobayashi}. This state 
consists of chain of spins ordered ferromagnetically in one direction and
antiferromagnetically in the perpendicular direction~\cite{yildirim, rotter,zhao,zhao1}. According 
to the conventional double-exchange mechanism, it was expected that the conductivity would be
better along the ferromagnetic direction~\cite{zener}. On the contrary, the transport measurements 
show that the conductivity is comparatively higher along the antiferromagnetic 
direction~\cite{chu, tanatar, ying}. The origin of this unusual behavior
has been attributed to the complex electronic structure resulting from the reconstruction in the SDW state~\cite{bascones1,daghofer,dheeraj1,dheeraj2}.

Apart from transport properties, there are several other 
experimental evidences which confirm the anisotropic electronic behavior in the four-fold rotation symmetry 
broken state of both the SDW state and the nematic state marked by non degenerate $d_{xz}$ and $d_{yz}$ orbitals. 
These experiments are angle-resolved photoemission spectroscopy (ARPES)~\cite{yi, shimojima1}, nuclear magnetic 
resonance (NMR)~\cite{imai}, magneto-torque measurements~\cite{kasahara}, scanning tunneling 
microscopy (STM)~\cite{chuang,dheeraj3,dheeraj4} etc. For instance, ARPES measurements shows the existence of orbital 
splitting between $d_{xz}$ and $d_{yz}$ orbitals below tetragonal-to-orthorhombic transition. The 
transition may occur above the SDW transition~\cite{nandi} or simultaneously~\cite{rotter1}. The Fermi pockets 
are mainly dominated by the $d_{xz}$ orbitals~\cite{shimojima}. The optical conductivity also shows a 
significant anisotropy~\cite{nakajima}.

As a result of a tetrahedral environment and a small distortion, the degeneracy of iron 3$d$ orbitals 
is partially lifted. However, the order of splitting can be as large as $\sim $ 0.5eV, which is small in comparison to the bandwidth. Therefore, all 3$d$ orbitals should be
included in any theoretical model for a realistic description of the properties of iron-based superconductors,
an important consequence of which is a complex bandstructure~\cite{kuroki,graser}. Theoretical studies of 
conductivity in the SDW state are mainly based on the mean-field~\cite{sugimoto,zhang,yin1} approaches which 
attribute the anisotropy to the difference in $d_{xz}$ and $d_{yz}$ orbital weights along the reconstructed Fermi surfaces dominated mainly by three $d_{xz}$, $d_{yz}$ and $d_{xy}$
orbitals. The role of $d_{xy}$ orbital, in particular, the interorbital hopping from $d_{xz}$ to $d_{xy}$ orbital has also been emphasized~\cite{yin1,zhang}. This behavior is shown 
to be robust against the change in the carrier concentration~\cite{dheeraj1}.  

On site Coulomb interactions are the additional factors which may also affect the 
conductivity anisotropy. Hund's coupling ($J$) is known to suppress the double occupancy 
through spin polarization~\cite{fanfarillo}. It may also play a significant role in orbital decoupling~\cite{bascones} and
orbital-selective Mott transition~\cite{yu,medici}. In the iron-based superconductors, various theoretical and 
experimental estimates suggest $J \sim  0.15U$ on the lower side~\cite{miyake} while $J \sim 0.25U$ on the 
higher side~\cite{ishida}. Therefore, an important question arises with regards to the nature of conductivity anisotropy when $J$ varies within the range $0.15U \lesssim J \lesssim 0.25U$. Furthermore, the role of $J$ in controlling the ferro-orbital order in the SDW state without four-fold rotation symmetry is also not clear, though it may suppress the same in the nematic state as indicated recently~\cite{laura}.

\begin{figure*}
    \centering
    \includegraphics[scale=1.0, width=15cm]{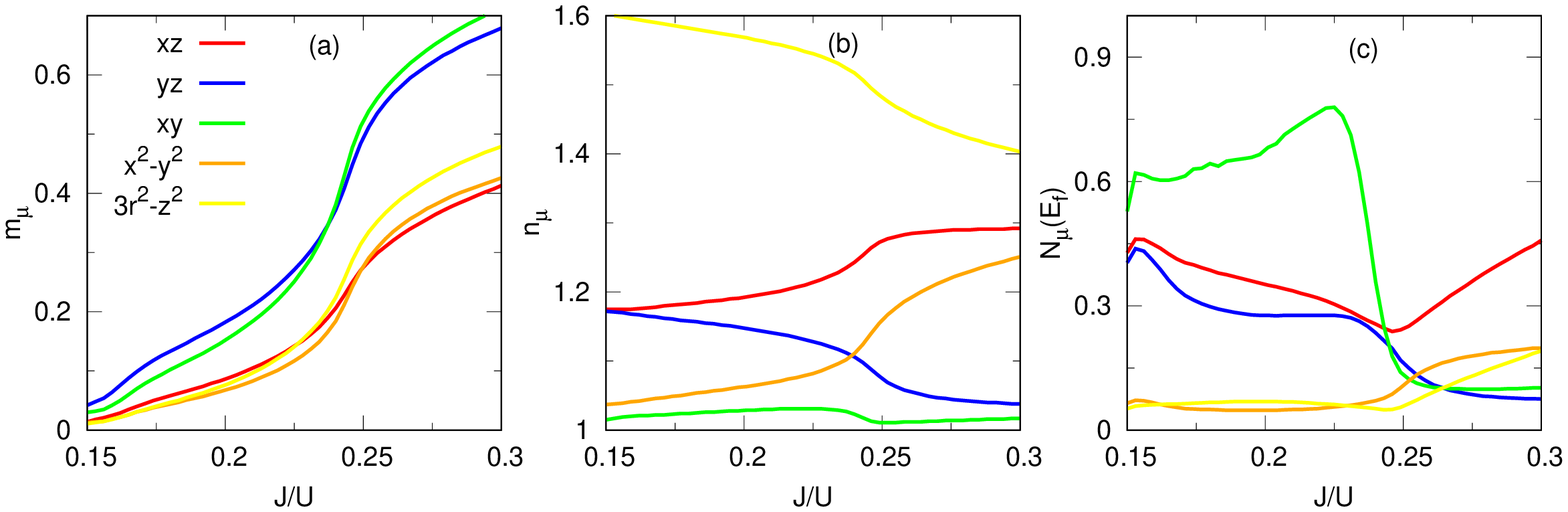}
    \vspace{0mm}
    \caption{(a) Orbital magnetizations, (b) orbital charge densities and (c) DOS at the Fermi level in ($\pi$,0) SDW state as a function of Hund's coupling for $U = 1.5$eV.}
    \label{1}
\end{figure*} 
In this paper, we examine the role of $J$ in Drude-weight anisotropy in the collinear ($\pi, 0$) SDW state, which is obtained self-consistently using the mean-field decoupling of on site Coulombic terms within a five-orbital model. In the ($\pi, 0$) SDW state, the spins have ferromagnetic and antiferromagnetic arrangements along $y$ and $x$ direction, respectively. We find (i) a sharp rise in the magnetization as a function of $J$ especially near $J \approx 0.25U$, (ii) an increase in ferro-orbital order with $0.15U \lesssim J \lesssim 0.25U$ beyond which it becomes nearly constant and (iii) dominant contribution of $d_{x^2-y^2}$ and $d_{3z^2-r^2}$ orbitals at the Fermi level near and beyond $0.25U \lesssim J$. (iv) We also find that the conductivity anisotropy is nearly robust in the close vicinity of $J \sim 0.25U$ despite a significant modification in the reconstructed Fermi surface topology and orbital-weight distribution all along. 
\section{Model and method}
In order to study the role of Hund's coupling in the conductivity anisotropy, we consider a multiorbital model with all the five 3$d$ orbitals~\cite{graser}. The kinetic part of the model Hamiltonian is 
\begin{equation}
    \mathcal{H}_0  = - \sum_{\bf i,j}^{\mu\nu \sigma} \textit{t}_{\bf i,j}^{\mu \nu} \textit{d}^{\dagger}_{{\bf i} l \sigma} \textit{d}^{}_{{\bf j} m \sigma} ,
\end{equation}
where $\textit{d}^{\dagger}_{{\bf i} \mu \sigma} (\textit{d}^{}_{{\bf i} \mu \sigma})$ is the electron creation (destruction) operator for an electron in $\mu$-th orbital with spin $\sigma$ at site {\bf i}. $\textit{t}_{\bf{i,j}}^{\mu \nu}$ are the hopping amplitudes from the orbital $\mu$ and $\nu$ of sites {\bf i} and {\bf j}, respectively. The orbitals $\mu$ and $\nu$ may denote any of the five 3$d$ orbitals $\textit{d}_{xz}$, $\textit{d}_{yz}$, $\textit{d}_{xy}$, $\textit{d}_{x^2 - y^2}$ and $\textit{d}_{3z^2 - r^2}$. The summation is over lattice sites, five orbital and spin states.

On site electron-electron interaction is given by
\begin{eqnarray}
\mathcal{H}_{int} &=& \mathcal{H}^{i}+\mathcal{H}^{d-d}+\mathcal{H}^{h}+\mathcal{H}^{p-h}\nonumber\\
&=& \textit{U} \sum_{{\bf i} \mu} \textit{n}_{{\bf i} \mu \uparrow} \textit{n}_{{\bf i} \mu \downarrow} + (\textit{U}^{\prime} - \frac{\textit{J}}{2}) \sum_{{\bf i}, \mu<\nu, \sigma {\sigma}^{\prime}} \textit{n}_{{\bf i} \mu \sigma} \textit{n}_{{\bf i} \nu {\sigma}^{\prime}} \nonumber\\ &-& 
    2 \textit{J} \sum_{{\bf i}}^{\mu < \nu} {\bf S}_{{\bf i} \mu} \cdot {\bf S}_{{\bf i} \nu} + \textit{J} \sum_{{\bf i} \sigma}^{\mu < \nu} \textit{d}_{{\bf i} \mu \sigma}^{\dagger} \textit{d}_{{\bf i} \mu \bar{\sigma}}^{\dagger} \textit{d}_{{\bf i} \nu \bar{\sigma}} \textit{d}_{{\bf i} \nu \sigma}. 
\end{eqnarray}    
First term $\mathcal{H}^{i}$ describes the Coulomb repulsion between the electrons of opposite spin but same orbital. Second term $\mathcal{H}^{d-d}$ accounts for the interorbital density-density interaction. $\mathcal{H}^{h}$ is the Hund's interaction for the electrons in different orbitals. Whereas $\mathcal{H}^{p-h}$ corresponds to the pair-hopping energy. We choose ${\textit{U}}^{\prime}$ =  $\textit{U} - 2 \textit{J}$ so that the interaction is rotation invariant~\cite{castellani}. In order to examine the role of $J$, $U$ is fixed to be 1.5eV throughout unless mentioned otherwise, which is in accordance with the estimates $U/W$ $\sim$ 0.25 of the intraorbital Coulomb interaction ($U$) by various studies, where the bandwidth is $W\sim6$eV~\cite{yang}.

In the following, we describe the interaction terms in the mean-field approximation, which consists of bilinear terms in the electron creation (or annihilation) operator in the SDW state. The mean-field decoupled terms, when the magnetic moments are oriented along $z$ direction, are given as 
\begin{equation}
\mathcal{H}_{\rm mf}^{i} =\frac{U}{2} \sum_{{\bf k} \mu \sigma} \bigg(- s \sigma  m_{\mu}  + n_{\mu}\bigg) d_{{\bf k}\mu\sigma}^\dagger d_{{\bf k}\mu\sigma},
\end{equation}
\begin{eqnarray}
\mathcal{H}^{d-d}_{\rm mf} = (U' - \frac{J}{2}) \sum_{{\bf k}, \mu \neq \nu, \sigma}  n_{\nu}  d_{{\bf k}\mu\sigma}^\dagger d_{{\bf k}\mu\sigma} 
\end{eqnarray}
and 
\begin{eqnarray}
\mathcal{H}^{h}_{\rm mf} = - \frac{J}{2} \sum_{{\bf k}, \mu \neq \nu, \sigma} \sigma s m_{\nu} d_{{\bf k}\mu\sigma}^\dagger d_{{\bf k}\mu\sigma}.
\end{eqnarray}
$n_{\mu}$ and $m_{\mu}$ are the charge density and sublattice magnetization for the orbital $\mu$, respectively. \textit{s} and $\sigma$ take value 1 (-1) for A (B) sublattice and $\uparrow$ spin ($\downarrow$ spin), respectively. For simplicity, we ignore the off-diagonal terms resulting from various terms including the pair-hopping as the off-diagonal terms in magnetization are usually very small in the absence of spin-orbit coupling.

The meanfield Hamiltonian ${H}_{mf}$ in the ($\pi$, 0) SDW state with two sublattice basis is \\
 \begin{equation}
     \mathcal{H}_{mf} = \sum_{{\bf k} \sigma} \Psi_{{\bf k} \sigma}^{\dagger} (\hat{t}_{{\bf k} \sigma} + \hat{M}_{{\bf k} \sigma}) \Psi_{{\bf k} \sigma},
 \end{equation}
 where matrix elements $\textit{t}_{{\bf k} \sigma}^{\mu \nu}$ and $\textit{M}_{{\bf k} \sigma}^{\mu \nu} = -s \sigma \Delta_{\mu \nu} + \frac{5 \textit{J} - \textit{U}}{2} \textit{n}_{\mu} \delta^{\mu \nu}$ are due to kinetic and interaction parts, respectively. $\mu$, $\nu \in$ \textit{s} $\otimes \, \tau$ with \textit{s} and $\tau$ denoting a sublattice and an orbital, respectively. The electron-field operator and exchange fields are defined as $\Psi_{k}^{\dagger} = ( \textit{d}_{A {\bf k}1 \uparrow}^{\dagger}, \textit{d}_{A {\bf k}2 \uparrow}^{\dagger}, \cdots \textit{d}_{B {\bf k}1 \uparrow}^{\dagger}, \textit{d}_{B {\bf k}1 \uparrow}^{\dagger},  \cdots )$ and $2 \Delta_{\mu \nu} = \textit{U m}_{\nu} \delta^{\mu \nu}  + \textit{J} \delta^{\mu \nu} \sum_{\mu \neq r} \textit{m}_{r}$, respectively. Charge density $\textit{n}_{\mu}$ and magnetization $\textit{m}_{\mu}$ in each of the orbitals are determined self-consistently by diagonalizing the Hamiltonian.

The optical conductivities $\sigma_{\alpha}$ along $\alpha = x \,\,{\rm or}\,\, y$ are calculated as~\cite{dagotto,valenzuela}  
\begin{eqnarray} \sigma_{\mathcal{\alpha}} &=& D_{\alpha} \delta{(\omega)}+
\frac {1}{N} \sum_{{\bf k}, n \ne n^{\prime}} 
\frac{ |j^{\alpha}_{nn^{\prime}}({\bf k})|^2}{E_{n^\prime {\bf k}} - 
E_{n {\bf k}}} \nonumber\\
&\times&\theta (-E_{n^\prime {\bf k}})\theta (E_{n {\bf k}}) 
\delta(\omega-E_{n^\prime {\bf k}}+E_{n {\bf k}}), 
\end{eqnarray}
where $D_{\alpha}$ is the Drude weight component with $\alpha = x,y$.
\begin{eqnarray} \frac {D_{\alpha}}{2 \pi} &=& \frac{1}{2N}
\sum_{\bf k} T^{\alpha}_{nn} ({\bf k}) \theta (-E_{n {\bf k}}) -
\frac {1}{N} \sum_{{\bf k}, n \ne n^{\prime}} 
\frac{ |j^{\alpha}_{nn^{\prime}}({\bf k})|^2}{E_{n^\prime {\bf k}} - 
E_{n {\bf k}}} \nonumber\\
&\times&\theta (-E_{n^\prime {\bf k}})\theta (E_{n {\bf k}}).
\end{eqnarray}
Here, in the SDW state, $E_{n{\bf k}}$ is the single particle energy, $\theta$ is the step function and the subscript $n (n^{\prime})$ denotes the band.
\begin{eqnarray} T^{\alpha}_{nn} = \sum_{\mu \nu} T^{\alpha;\mu \nu}_{nn}
= \sum_{\mu \nu} \frac {\partial^2 
t^{\mu \nu} ({\bf k})}{\partial k^2_x}
a^*_{{\bf k} \mu n} a_{{\bf k}\nu n} 
\nonumber\\
j^{\mathcal{\alpha}}_{nn^{\prime}} = -\sum_{\mu \nu} j^{\alpha;\mu \nu}_{nn^{\prime}} 
= -\sum_{\mu \nu} \frac {\partial
t^{\mu \nu} ({\bf k})}{\partial k_x} 
a^*_{{\bf k} \mu n} a_{{\bf k}\nu n},
\end{eqnarray}
$a_{{\bf k} \mu  n}$ is the unitary matrix element 
between the orbital $\mu$ and band $n$. Lorentzian is used to approximate the $\delta$ function, where a small but same broadening parameter is used in both directions.

Our focus in this work is the anisotropy in $\omega \rightarrow 0$ limit. In order to gain a better insight into the link between the orbital weight redistribution and anisotropy, we also examine Drude weight components along different directions as below~\cite{zhang}.

\section{Result and discussion}

 Fig.~\ref{1} (a) shows the orbital-resolved magnetizations as a function of $J$ with $U =$ 1.5eV. The magnetization in each orbital increases monotonically with $J$ as expected because the Hund's interaction is known to maximize the total spin. Noticeably, there is a steep rise in the magnetization near \textit{J} $\sim 0.25U$, which may be indicative of an enhanced stabilization of the SDW state for the intermediate value of Hund's coupling. On the other hand, orbital magnetization becomes very small below $J \lesssim 0.15U$, which may correspond to a rather fragile SDW state. Throughout the range $0.15U \lesssim J \lesssim 0.25U$, the dominant contribution to the net magnetization comes mainly from the $d_{xz}$ and $d_{xy}$ orbitals with nearly equal magnetic moments. 
 Their individual contribution is almost double of the contributions made by any of the remaining orbitals. This feature can be understood with the help of orbital occupancies. With a rise in $J$, $d_{yz}$ and $d_{xy}$ orbitals approach half filling for the reasons to be discussed below, which leads to large magnetic moments of these orbitals.

 \begin{table*}[t]
\caption{Elements of Drude weight for $U = 1.5$eV. Orbital indices having values 1, 2, 3, 4, and 5, correspond to the  $d_{xz}$, $d_{yz}$, 
$d_{x^2-y^2}$, $d_{xy}$, and $d_{3z^2-r^2}$ orbitals, respectively. Subscript $i$ = 1, 2 and 3 in $D^{\alpha \beta}_{x;i}$ correspond to three different cases $J = 0.15U$, $J = 0.225U$ and $J = 0.3U$.} 
 \vspace{5mm}
\centering 
\setlength{\tabcolsep}{2.5pt} 
    \renewcommand{\arraystretch}{1} 
\begin{tabular}{c|c c c c c c c c c c c c c c c c} 
\hline\hline 
$\alpha \beta$ & $11$ & $14$ & $22$ &  $24$ & $35$ & $44$ & $45$ & $55$\\ [0.5ex] 
\hline
$D^{\alpha \beta}_{x;1}$ & $0.0936$ &  $0.2608$  & $0.0246$  &  $0.000$ & $-0.0003$ & $0.0619$ &  $0.0027$& $0.000$\\
\hline
$D^{\alpha \beta}_{y;1}$ & $ 0.0194$&  $0.0093$  & $0.0564$  &  $0.1491$  & $-0.0004$ & $0.2223$ & $0.0010$& $0.000$\\
\hline
$D^{\alpha \beta}_{x;2}$ & $0.0559$ &  $0.1248$  & $0.0129$  &  $0.0001 $ & $0.0026$ & $0.0643$ &  $0.0009$& $0.000$  \\
\hline
$D^{\alpha \beta}_{y;2}$ & $0.0313$ &  $0.0003$ & $0.0013$  &  $0.0155$  & $0.0006$ &  $0.0564$ &  $0.0054$& $0.00$ \\
\hline
$D^{\alpha \beta}_{x;3}$ & $0.0202$   & $0.0790$  & $0.0008$ & $ 0.0001$ & $0.0353$& $ 0.0522$ & $ 0.0024$& $-0.0039$\\
\hline
$D^{\alpha \beta}_{y;3}$ & $0.0238$   & $-0.0002$ & $0.0038$ & $ 0.0150$ &$0.0178$ & $ 0.0003$ & $ 0.0285$& $0.0115$\\    
[1ex] 
\hline \hline
\end{tabular}
\end{table*} 

 Fig.~\ref{1} (b) shows the orbital-resolved charge density as a function of $J$. With increase in $J$, the occupancy of $d_{xz}$ and $d_{x^2 - y^2}$ orbitals increases while that of the remaining orbitals decreases monotonically. Further, $n_{xz} \approx n_{yz}$ and $n_{xy} \approx n_{x^2-y^2}$ for smaller values of $J$. The ferro-orbital order associated with four-fold rotation symmetry and defined as $n_{xz} - n_{yz}$ continues to increase with $J$ exhibiting a steep rise near $J \sim 0.25U$ and becomes constant thereafter. A similar feature is noticed for the difference $n_{x^2-y^2} - n_{xy}$. Therefore, the Hund's coupling doesn't only stabilize the magnetic order but it may also help in enhancing the orbital order for the realistic range $ 0.15U \lesssim J \lesssim 0.25U$. 
 
 Aforementioned behavior of the orbital-charge densities and particularly of the ferro-orbital order results from the interplay between three important factors, which are the on-site Coulomb repulsion, the crystal-fields splitting and the nature of existent magnetic order. A strong Coulomb repulsion will minimize the double occupancy. Further, it may also be noted that the double occupancy of a low-lying orbitals like $d_{3z^2-r^2}$ is preferable in comparison to the high-lying $d_{xz}$ or $d_{yz}$ orbitals as it leads to energy reduction. On the other hands, collinear magnetic order with broken four-fold rotation symmetry yields an unequal occupancy of $d_{xz}$ and $d_{yz}$ otherwise degenerate orbitals.
 
\begin{figure}
       \includegraphics[scale=1.0, width=8.8cm]{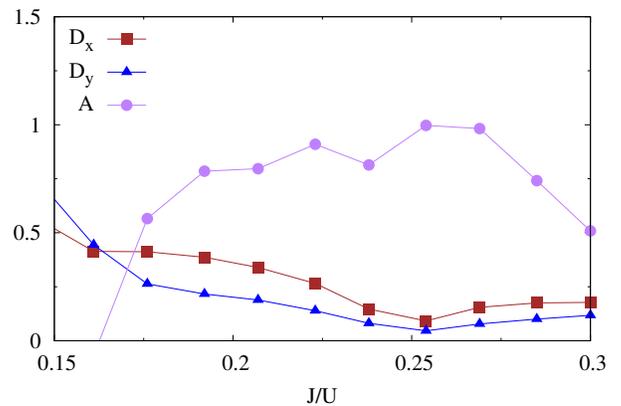}
    \hspace{-0mm}
    \vspace{-5mm}
    \caption{Drude weights $D_y$ and $D_x$ along the ferromagnetic and antiferromagnetic directions as a function of $J$ with $U$ = 1.5eV, respectively along with $A$ = $\frac{D_x}{D_y}-1$, which is a measure of anisotropy. }
    \label{2}
\end{figure} 
\begin{figure*}
\hspace{0cm}
\includegraphics[scale=1.0, width=15.00cm]{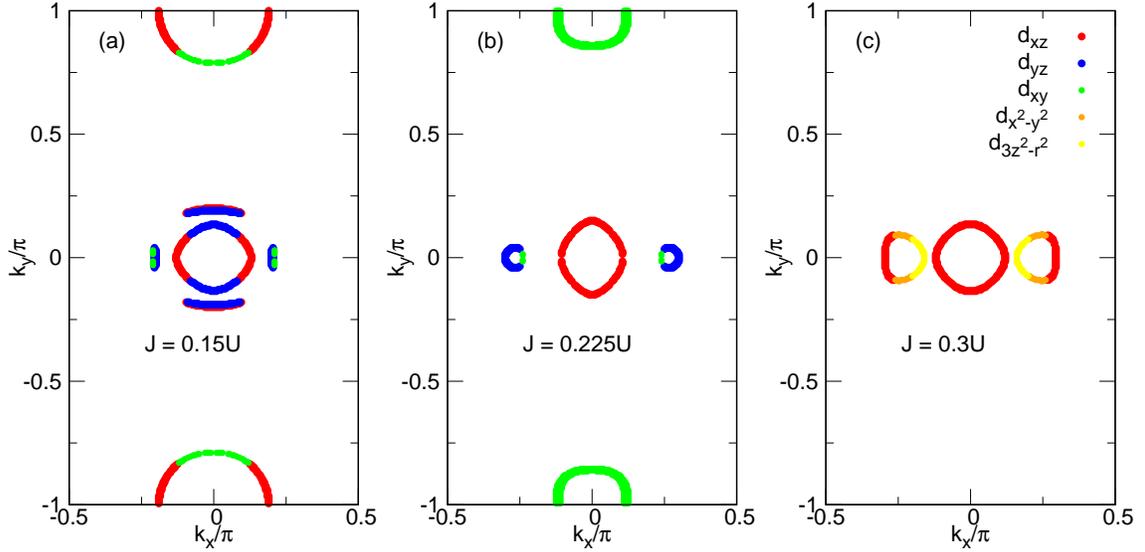}
     \vspace{-5mm}
\caption{Fermi surfaces for different Hund's coupling $J$ = (a) $0.15U$, (b) $0.225U$ and (c) $0.30U$ with $U = 1.5$eV. Different colors have been used to show the orbitals predominant along the surfaces.}
\label{3}
\end{figure*}

In order to the understand the rise of ferro-orbital order parameter with $J$, we first note that the nearest-neighbor hopping parameters in the model of Graser \textit{et. al.} for the $d_{xz}$ and $d_{yz}$ orbitals are $t_{1x} = 0.4$eV and $t_{2x} = 0.14$eV, where the subscripts 1 and 2 correspond to $d_{xz}$ and $d_{yz}$ orbitals, respectively \cite{graser}. Initially, when $J \sim 0.15U$eV, $n_{xz} - n_{yz} \sim 0$, then it grows thereafter with an increase in $J$. This enhancement is accompanied with a rise and drop in electron fillings of $d_{xz}$ and $d_{yz}$ orbitals, respectively. For a given $U$, a large $J$ enhances the total magnetic moment by maximizing the spin-polarization of the orbitals. With a rise in the magnetic moments, the difference in the exchange field experienced by an $\uparrow$-spin electron hopping from one site with magnetization along $+z$ to another neighboring site along the antiferromagnetic direction, \textit{i.e.}, $2 \Delta_{\mu \mu} = \textit{U m}_{\mu}  + \textit{J} \sum_{r \neq \mu} \textit{m}_{r}$ increases. As a result, the energy cost of the double occupancy of orbitals also increases, which is expected to lead to localization of electrons for an orbital with a larger $U/t_x$. 

In the SDW state, the energy can be minimized by delocalization energy gain along $x$ direction through an orbital which has a relatively smaller $U/t_x$ ratio and a filling greater than half filling. For $d_{xz}$ orbital $U/t_{2x} \sim 4$, thus, an $\uparrow$-spin electron from a site with magnetic moment aligned along negative $z$ direction can hop to a nearby site with less energy cost as compared to $\downarrow$-spin electron. Secondly, $U/t_{2x} \gtrsim 10$ for $d_{yz}$ orbital is already within the strong coupling regime and the electron in this orbital find it difficult to hop along $x$ direction with a large energy cost for double occupancies, a scenario which supports a half-filled $d_{yz}$ orbital. However, the second-order exchange process can take place with $J_e \sim (t_{2x})^2/U$, which may help in stabilizing the antiferromagnetic spin-arrangement along the $x$ direction. Thus, a larger occupancy of $d_{xz}$ orbital and hence the ferro-orbital order may be seen as a consequence of the antiferromagnetic spin arrangement along $x$ direction which is supported by a half-filled $d_{yz}$ orbital instead. 

Because of half filling, $d_{yz}$ electrons find it difficult to move along the ferromagnetic direction through the intraorbital hopping process as the $\uparrow$-spin state is already occupied in the neighboring sites. The same is true for the interorbital hopping process, $d_{yz}$ electrons cannot hop to neighboring $d_{xz}$ orbitals, because $\uparrow$-spin $d_{xz}$ state is also occupied. Thus $d_{yz}$ electrons become increasingly localized. 

Beyond $J \sim 0.25U$, the orbital order remains constant as $n_{xz}$ does not show any significant change. In particular, it doesn't rise because the magnetization $d_{yz}$ orbital magnetization in that region is not far from it's asymptotic value and there is neither significant change in energy cost of double occupancy of $d_{yz}$ orbitals nor delocalization energy gain by increasing $d_{xz}$ orbital filling.

In addition to $d_{yz}$, another nearly half-filled orbital is $d_{xy}$. As can be noticed from Fig. (\ref{1}), $n_{xy} \sim 1$ for all values of $J$. An almost half-filled $d_{xy}$ orbital for small $J$ is mainly a consequence of it's relative location with respect to the crystal filled splitting. However, the nearest-neighbor hopping parameter $t_{3x} = 0.23$eV for $d_{xy}$ orbital indicates $U/t_{3x} \sim 7$, a value towards the boundary of intermediate to strong coupling regime, which ensures that $n_{xy} \sim 1$ for even larger $J$ for the reason discussed above.  

As noted earlier, on approaching $J \sim 0.25U$, there is a sharp rise in the ferro-orbital order, which is accompanied with $d_{yz}$ and $d_{xy}$ orbitals approaching half filling. Consequently, there is also a sharp increase in the magnetization in these two orbitals, which further drives a steep rise in the magnetization of the remaining orbitals through Hund's coupling. 
 
Fig.~\ref{1} (c) shows the orbital weight at the Fermi level. For smaller $J$, $d_{xy}$ orbital is predominant along the Fermi surfaces. Other dominating orbitals are $d_{xz}$ and $d_{yz}$. $d_{yz}$ contribution decreases monotonically with an increase in $J$ throughout the range considered. $d_{xy}$ orbital weight increases and maximizes near $J \sim 0.225U$, it shows a sharp decline afterwards. Whereas $d_{xz}$ orbital contribution first declines with increasing $J$, minimizes near $J \sim 0.225U$ and increases thereafter. The orbital weight of the remaining orbitals are small. However, beyond $J \sim 0.25U$, the three dominating orbitals at the Fermi surface are $d_{xz}$, $d_{x^2-y^2}$ and $d_{3r^2-z^2}$ orbitals instead of the set $d_{xz}$, $d_{yz}$ and $d_{xy}$ orbitals. Thus, we find that the orbital-weights redistribution along the reconstructed Fermi pockets are very sensitive to the Hund's coupling.     

Fig.~\ref{2} shows the Drude weights and anisotropy as a function of $J$. For a smaller $J \sim 0.15U$, the Drude weights ($D_{x}$ and $D_{y}$) have large values because of the presence of several Fermi pockets. This is a consequence of a weak reconstruction of electronic bandstructure in the SDW state for small $J$. With an increase in $J$, both $D_{x}$ and $D_{y}$ decrease continuously as $J$ approaches $0.25U$, which results from the disappearance of several Fermi pockets. Both $D_{x}$ and $D_{y}$ show a minor increase beyond $J \sim 0.25U$ resulting from the enlargement of the smaller hole pockets adjacent to the largest but robust hole pocket centered around $\Gamma$. However, the parameter $A$, which can be used to quantify the anisotropy, defined below 
\begin{equation}
A = \frac{D_{x}}{D_{y}} - 1 
\end{equation}
is robust and doesn't change much for the intermediate value of $J$. Interestingly, the ferromagnetic direction conducts better for a smaller value of $J$ as evident from a negative $A$, which   results from a weak electronic reconstruction.

Characteristics of the Drude weight as mentioned above can be understood in terms of Fermi velocities oriented in a direction perpendicular to the Fermi surfaces and the anisotropic orbital-weight redistribution. Fig.~\ref{3} shows the Fermi surfaces along with the dominating orbitals. We first note that for $U = 1.5$eV, several features of the Fermi surfaces for $J \sim 0.25U$ are found to be in agreement with ARPES measurement. These features include the hole pocket around $\Gamma$ and electron pockets located along  $\Gamma$-X and $\Gamma$-Y directions, respectively~\cite{yi1}.

As anticipated from the dominating orbitals along the Fermi surface, main contributors to $D_{x}$ for smaller $J \sim 0.15U$ are $D^{11}$, $D^{14}$ and $D^{44}$ whereas $D^{24}$ and $D^{44}$ dominate the contribution to $D_{y}$ (Table 1). Highly anisotropic contribution of $D^{11}_x$ is because of $v^f_x > v^f_y$ in the region of electron pockets centered around $(0, \pm \pi)$ dominated by $d_{xz}$ orbital. The same electron pocket is responsible for a large $D^{14}_x$ but vanishingly small $D^{14}_y$. However, the net anisotropy is small but negative, \textit{i.e}, ferromagnetic direction conducts better. This is primarily because of $D^{24}$ originating from tiny electron pockets around $(\pm 0.2\pi, 0)$ and $D^{44}$ from the electron pocket around ($\pi, 0$) contributing mainly to $D_{y}$. It may be noted that  $v^f_y > v^f_x$ in the regions dominated by $d_{xy}$ orbital. When $J$ is increased further, the anisotropy in $D^{11}$ declines whereas that of $D^{44}$ is reversed. The latter is a consequence of the electron pockets around $(0, \pm \pi)$, which becomes completely dominated by $d_{xy}$ orbital. Also $D^{24}$ becomes negligibly small in both directions, so that there is an overall increase in anisotropy. On further increasing $J$, the anisotropy decreases as $D^{14}_x$ and $D^{44}_x$ both decrease while  $D^{45}_y$ and $D^{55}_y$ become non negligible. In this way, $A$ increases with $J$, maximizes near $J = 0.25U$ with it's value $A \approx 1$ and then decreases further. 

\section{summary and conclusions}
To conclude, we have investigated the role of Hund's coupling on  the shape and orbital content of reconstructed Fermi surfaces and how it impacts the conductivity anisotropy. The detailed analysis of Drude weights along different direction in terms of orbital weight indicates most significant role of intraorbital hopping between $d_{xy}$ orbitals as well as interorbital hopping from $d_{xy}$ to $d_{xz}$ and $d_{yz}$ orbitals for smaller $J\sim0.15U$ while a dominant role of inter-orbital hopping from $d_{xy}$ to $d_{xz}$ orbital up to $J\sim0.25U$. In the vicinity of $J\sim0.25U$, the remaining orbitals also begin to contribute.  Beyond $J\sim0.25U$, the intraorbital hopping between $d_{3z^2-r^2}$ and interorbital hopping between $d_{x^2-y^2}$ and $d_{3z^2-r^2}$ orbitals contribute significantly, which results in the reduction of anisotropy. 

The ferro-orbital order is found to increase with $J$ for a given $U$ in the widely considered regime $0.15U \lesssim J \lesssim 0.25U$. This may be seen as a consequence of increased double occupancy cost enabled by a rise in magnetization with $J$. It leads to an increased occupancy of $d_{xz}$ orbital away from half filling and depletion of $d_{yz}$ orbital towards half filling, which may stabilize the antiferromagnetic arrangement along $x$ direction through the second-order exchange process.

\acknowledgments
DKS was supported through start-up research grant SRG/2020/002144 funded by DST-SERB, India.


\begin{thebibliography}{0}



 \bibitem{lee} C.-C. Lee, W.-G. Yin and W. Ku, Phys. Rev. Lett. {\bf 103}, 267001 (2009).
 
  \bibitem{johnston} D. C. Johnston, Adv. Phys. {\bf 59}, 803 (2010).
 
 \bibitem{stewart} G. R. Stewart, Rev. Mod. Phys. {\bf 83}, 1589 (2011).
 
 \bibitem{dai} P. Dai, J. Hu and E. Dagotto, Nat. Phys. {\bf 8}, 709 (2012).
 
 \bibitem{shimojima} T. Shimojima, K. Ishizaka, Y. Ishida, N. Katayama, K. Ohgushi, T. Kiss, M. Okawa,
  T. Togashi, X.-Y. Wang, C.-T. Chen, S. Watanabe, R. Kadota, T. Oguchi,
  A. Chainani and S. Shin, Phys. Rev. Lett. {\bf 104}, 057002 (2010).

\bibitem{greve}  J. H. Chu, J.-G. Analytis, K. De Greve, P. L. McMahon,
Z. Islam, Y. Yamamoto and I. R. Fisher, Science {\bf 329}, 824  (2010).

\bibitem{liang} S. Liang, G. Alvarez, C. Sen, A. Moreo and E. Dagotto, Phys. Rev. Lett. {\bf 109}, 047001 (2012).

\bibitem{fernandes} R. M. Fernandes, A. V. Chubukov, J. Knolle, I. Eremin and J. Schmalian, Phys. Rev. B. {\bf 85}, 024534 (2012).

\bibitem{fernandes1} R. M. Fernandes, A. V. Chubukov and J. Schamalian, Nat. Phys. {\bf 10}, 97 (2014).
 
\bibitem{blomberg} E. C. Blomberg, M. A. Tanatar, R. M. Fernandes, I. I. Mazin, B. Shen, H.-W. Wen, M. D. Johannes, J. Schmalian and R. Prozorov, Nat. Commun. {\bf 4}, 1914 (2013). 


\bibitem{kobayashi} T. Kobayashi, K. Tanaka, S. Miyasaka and S. Tajima, J. Phys.
Soc. Jpn. {\bf 84}, 094707 (2015). 

 \bibitem{yildirim} T. Yildirim, Phys. Rev. Lett., {\bf 101}, 057010 (2008). 
 
 \bibitem{rotter} M. Rotter, Phys. Rev. B. {\bf 78}, 020503(R) (2008).
 
 \bibitem{zhao} J. Zhao, Q. Huang, C. de la Cruz, S. Li, J. W. Lynn, Y. Chen, 
 M. A. Green, G. F. Chen, G. Li, Z. Li, J. L. Luo, N. L. Wang and
 P. Dai, Nat. Mat. {\bf 7}, 953 (2008). 

\bibitem{zhao1}
 J. Zhao, D. T. Adroja, D. -O. Yao, R. Bewley, S. Li, X. F. Wang,
 G. Wu, X. H. Chen, J. Hu and P. Dai, Nat. Phys. {\bf 5}, 555 (2009). 

\bibitem{zener} C. Zener, Phys. Rev. {\bf 82}, 403 (1951).

 \bibitem{chu} J.-H. Chu, J.-H. Chu, J. G. Analytis, K. De Greve, P. L.
 McMahon, Z. Islam, Y. Yamamoto, and I. R. Fisher, Science
 {\bf 329}, 824 (2010).
 
 \bibitem{tanatar} M. A. Tanatar, E. C. Blomberg, A. Kreyssig, M. G. Kim, 
  N. Ni, A. Thaler, S. L. Bud'ko, P. C. Canfield, A. I. Goldman, I. I. Mazin,
 and R. Prozorov, Phys. Rev. B. {\bf 81}, 184508 (2010).

\bibitem{ying} J. J. Ying, X. F. Wang, T. Wu, Z. J. Xiang, R. H. Liu, 
 Y. J. Yan, A. F. Wang, M. Zhang, G. J. Ye, P. Cheng, J. P. Hu and 
 X. H. Chen, Phys. Rev. Lett. {\bf 107}, 067001 (2011).

\bibitem{bascones1} B. Valenzuela, E. Bascones, and M. J. Calder\'{o}n,
Phys. Rev. Lett. {\bf 105}, 207202 (2010).

\bibitem{daghofer}
 M. Daghofer, Q.-L. Luo, R. Yu, D. X. Yao, A. Moreo and E. Dagotto,
Phys. Rev. B {\bf 81}, 180514(R) (2010).

\bibitem{dheeraj1}
 D. K. Singh and P. Majumdar,
Phys. Rev. B {\bf 98}, 195130 (2018). 

\bibitem{dheeraj2}
D. K. Singh and Y. Bang, Phys. Rev. B {\bf 101}, 155101 (2020). 

\bibitem{yi} M. Yi, D. Lu, J.-H. Chu, J. G. Analytis, A. P. Sorini,
 A. F. Kemper, B. Moritz, S.-K. Mo, R. G. Moore, M. Hashimoto, W.-S. Lee, 
 Z. Hussain, T. P. Devereaux, I. R. Fisher and Z.-X. Shen, Proc. Natl. Acad. Sci., USA {\bf 108}, 6878 (2011).

\bibitem{shimojima1} T. Shimojima, T. Sonobe, W. Malaeb, K. Shinada, A. Chainani,
 S. Shin, T. Yoshida, S. Ideta, A. Fujimori, H. Kumigashira, K. Ono, Y. Nakashima, 
 H. Anzai, M. Arita, A. Ino, H. Namatame, M. Taniguchi, M. Nakajima, S. Uchida, Y. Tomioka, 
 T. Ito, K. Kihou, C. H. Lee, A. Iyo, H. Eisaki, K. Ohgushi, S. Kasahara, T. Terashima, H. Ikeda, 
 T. Shibauchi, Y. Matsuda and K. Ishizaka,
Phys. Rev. B. {\bf 89}, 045101 (2014). S. L. Bud’ko

\bibitem{imai} M. Fu, D. A. Torchetti, T. Imai, F. L. Ning, J.-Q. Yan, and A. -S. Sefat, Phys. Rev. Lett. {\bf 109}, 247001 (2012).

\bibitem{kasahara}
 S. Kasahara,H. J. Shi, K. Hashimoto, S. Tonegawa, Y. Mizukami, T. Shibauchi, K. Sugimoto, T. Fukuda, T. Terashima, 
 A. H. Nevidomskyy and Y. Matsuda, Nature {\bf 486}, 382 (2012).

\bibitem{chuang} T.-M. Chuang, M. P. Allan, J. Lee, Y. Xie, N. Ni, S. L. Bud’ko,
 G. S. Boebinger, P. C. Canfield and J. C. Davis, Science {\bf 327}, 181 (2010).
 
\bibitem{dheeraj3} D. K. Singh and P. Majumdar, Phys. Rev. B {\bf 96}, 235111 (2017).

\bibitem{dheeraj4} D. K. Singh, A. Akbari and P. Majumdar, Phys. Rev. B {\bf 98}, 180506(R) (2018). 

\bibitem{nandi}  S. Nandi, M. G. Kim, A. Kreyssig, R. M. Fernandes, D. K. Pratt,
 A. Thaler, N. Ni, S. L. Bud’ko, P. C. Canfield, J. Schmalian, R. J. McQueeney and 
 A. I. Goldman, Phys. Rev. Lett. {\bf 104}, 057006 (2010).

\bibitem{rotter1} M. Rotter, M. Tegel, D. Johrendt, I. Schellenberg, W. Hermes,
and R. Pöttgen, Phys. Rev. B. {\bf 78}, 020503(R) (2008).

\bibitem{nakajima} M. Nakajima, T. Liang, S. Ishida, Y. Tomioka, K. Kihou, 
 C. H. Lee, A. Iyo, H. Eisaki, T. Kakeshita, T. Ito, and S. Uchida, Proc.
Natl. Acad. Sci. {\bf 108}, 12238 (2011).

\bibitem{kuroki} K. Kuroki, S. Onari, R. Arita, H. Usui, Y. Tanaka, H. Kontani and H. Aoki, 
Phys. Rev. Lett. {\bf 101}, 087004 (2009). 

\bibitem{graser} S. Graser, T. Maier, P. Hirschfeld and D. Scalapino, New J. Phys. {\bf 11}, 025016 (2009). 
  

\bibitem{sugimoto} K. Sugimoto, E. Kaneshita and T. J. Tohyama, Phys. Soc. Jpn.
{\bf 80}, 033706 (2011). 

\bibitem{zhang} X. Zhang and E. Dagotto, Phys. Rev. B. {\bf 84}, 132505 (2011). 

\bibitem{yin1} Z. P. Yin, K. Haule and G. Kotliar, Nat. Phys. {\bf 7}, 294 (2010). 


\bibitem{fanfarillo} L. Fanfarillo and E. Bascones, Phys. Rev. B {\bf 92}, 075136 (2015).

\bibitem{bascones} E. Bascones, B. Valenzuela and M. J. Calder\'{o}n,
Phys. Rev. B {\bf 86}, 174508 (2012). 


\bibitem{yu} R. Yu and Q. Sai, Phys. Rev. B {\bf 86}, 085104 (2010). 

\bibitem{medici} L. de' Medici, G. Giovannetti, M. Capone, 
Phys. Rev. Lett. {\bf 112}, 177001 (2014). 

\bibitem{miyake} T. Miyake, K. Nakamura, R. Aarit and M., J. Imada Phys. Soc.
Jpn. {\bf 79}, 044705 (2010).


\bibitem{ishida} H. Ishida and A. Liebsch, Phys. Rev. B. {\bf 81}, 054513 (2010).

\bibitem{laura} L. Fanfarillo, G. Giovannetti, M. Capone and E. Bascones, Phys. Rev. B {\bf 95}, 144511 (2017).

 \bibitem{castellani} C. Castellani, C. R. Natoli and J. Ranninger, Phys. Rev. B. {\bf 18}, 5001 (1978).

 \bibitem{yang} W. L. Yang, A. P. Sorini, C-C. Chen, B. Moritz, W.-S. Lee, F. Vernay, P. Olalde-Velasco, J. D. Denlinger, B. Delley, J.-H. Chu, J. G. Analytis, I. R. Fisher, Z. A. Ren, J. Yang, W. Lu, Z. X. Zhao, J. van den Brink, Z. Hussain, Z.-X. Shen, and T. P. Devereaux, Phys. Rev. B {\bf 80}, 014508 (2009).

\bibitem{dagotto} E. Dagotto, Rev. Mod. Phys. {\bf 66}, 763 (1994).
 
 \bibitem{valenzuela} B. Valenzuela, M. J. Calder\'{o}n, G. Le\'{o}n, and E. Bascones, Phys. Rev. B. {\bf 87}, 075136 (2013).

\bibitem{yi1} M. Yi, D. H. Lu, J. G. Analytis, J.-H. Chu, S.-K. Mo, R.-H. He,
M. Hashimoto, R. G. Moore, I. I. Mazin, D. J. Singh, Z. Hussain,
I. R. Fisher and Z.-X. Shen, Phys. Rev. B {\bf 80}, 174510 (2009).
 

 \end{thebibliography}
  \end{document}